\documentclass{article}

\usepackage{arxiv}

\usepackage[utf8]{inputenc} 
\usepackage[T1]{fontenc}    
\usepackage{hyperref}       
\usepackage{url}            
\usepackage{booktabs}       
\usepackage{amsfonts}       
\usepackage{nicefrac}       
\usepackage{microtype}      
\usepackage{lipsum}
\usepackage{graphicx}
\graphicspath{ {./images/} }

\title{Envisioning Responsible Quantum Software Engineering and Quantum Artificial Intelligence}

\author{
 Muneera Bano \\
   CSIRO's Data61\\
  Australia \\
  \texttt{muneera.bano@csiro.au} \\
   \And
 Shaukat Ali \\
  Simula Research Laboratory \& \\
  Oslo Metropolitan University\\
  Oslo, Norway \\
  \texttt{shaukat@simula.no} \\
  \And
 Didar Zowghi \\
   CSIRO's Data61\\
  Australia \\
  \texttt{didar.zowghi@csiro.au} \\
}

\begin{document}
\maketitle
\begin{abstract}
The convergence of Quantum Computing (QC), Quantum Software Engineering (QSE), and Artificial Intelligence (AI) presents transformative opportunities across various domains. However, existing methodologies inadequately address the ethical, security, and governance challenges arising from this technological shift. This paper highlights the urgent need for interdisciplinary collaboration to embed ethical principles into the development of Quantum AI (QAI) and QSE, ensuring transparency, inclusivity, and equitable global access. Without proactive governance, there is a risk of deepening digital inequalities and consolidating power among a select few. We call on the software engineering community to actively shape a future where responsible QSE and QAI are foundational for ethical, accountable, and socially beneficial technological progress.
\end{abstract}


\section{Introduction}

Quantum computing (QC) is a rapidly evolving field with the potential to revolutionise computation through quantum mechanical principles such as superposition and entanglement~\cite{preskill2018quantum}. These capabilities enable quantum computers to perform complex calculations exponentially faster than classical computers, opening new frontiers in cryptography, optimisation, and scientific simulations~\cite{schuld2019quantum}. One of the most promising applications of QC is in artificial intelligence (AI), where Quantum AI (QAI) leverages quantum capabilities to enhance classical AI, accelerate optimisation tasks, and improve decision-making systems~\cite{wang2023recent, herman2023quantum, ho2024quantum}. The convergence of QC and AI has the potential to drive breakthroughs in areas such as drug discovery, financial modelling, and climate simulation. However, it also introduces significant challenges for software engineering (SE)~\cite{ali2022software, de2024quantum}. Existing SE principles, designed for classical deterministic computing, are ill-equipped to handle the inherent uncertainties and probabilistic nature of quantum computation, necessitating the development of Quantum Software Engineering (QSE) to ensure the secure, transparent, and ethical deployment of QAI applications~\cite{ali2022software, sarkar2024automated, hoffmann2024business}.

Despite its promised transformative potential, QAI raises substantial ethical and security concerns, particularly in the areas of bias, explainability, interpretability, and accountability. Classical AI models have already demonstrated tendencies to reinforce biases embedded in training data, leading to discriminatory outcomes in critical domains such as healthcare and recruitment~\cite{shams2023ai, bano2024diversity, bano2025does}. Given QAI's exponential computational power, these biases could become further entrenched and harder to detect, intensifying concerns about algorithmic fairness and transparency~\cite{sarkar2024automated}. Moreover, the inherent lack of explainability in QAI decisions complicates accountability in high-stakes applications such as criminal justice and financial systems~\cite{AIQCConvergence}. If even developers cannot fully interpret how QAI models reach their conclusions, ensuring trustworthy decision-making and ethical AI governance becomes increasingly complex.

Beyond technical concerns, QAI also risks exacerbating global inequalities due to restricted access to quantum infrastructure and expertise. QC development is highly resource-intensive, requiring specialised hardware and significant financial investment, making it accessible to only a handful of well-funded corporations and governments~\cite{seskir2023democratization}. This exclusivity raises the risk of technological monopolisation, where a select few entities dictate the trajectory of QAI research and development, leaving developing nations and underprivileged communities struggling to keep pace~\cite{ten2023reading}. Without proactive governance and equitable distribution of quantum resources, QAI could become a tool for economic dependence and digital colonialism, reinforcing existing global disparities~\cite{bcg2022quantumrace}. Furthermore, QAI’s ability to generate hyper-realistic deepfakes, manipulate information at an unprecedented scale, and challenge traditional cybersecurity frameworks heightens the urgency for robust ethical safeguards and international regulatory collaboration~\cite{AIQCConvergence, meyer2023question, meyer2024disparities}.

Given these risks, this paper explores the ethical implications of QAI from a software engineering perspective. It examines how responsible QSE can guide the ethical development and deployment of QAI, ensuring these technologies remain transparent, fair, and accessible, rather than amplifying existing societal and economic disparities.

The discussion is structured as follows: Section \ref{sec:QCandAI} provides an overview of QAI and its emerging capabilities, followed by an analysis of key ethical dilemmas in Section \ref{QSE}. Section \ref{sec:ethics} explores the role of responsible QSE in mitigating these risks.

\section{Quantum Computing and AI} \label{sec:QCandAI}

Quantum Computing (QC) leverages principles of quantum mechanics, such as superposition, entanglement, and quantum parallelism, to perform computations~\cite{AIQCConvergence}. Unlike classical computing, which relies on binary logic (bits represented as 0s and 1s), QC operates on quantum bits (qubits), which can exist in superposition states, enabling computational speedups for some problem domains.

One of the most well-known examples of QC’s potential is Shor’s algorithm, which can factor large integers exponentially faster than any known classical algorithm, posing a direct threat to current cryptographic systems~\cite{shor2002introduction}. Similarly, Grover’s search algorithm offers quadratic speedups for database searching, improving efficiency in unstructured search problems~\cite{rungta2009quadratic}. Despite these promising advancements, practical QC remains in its early stages. While major research institutions and technology companies such as IBM, Google, and Chalmers University (Sweden) have developed early quantum processors, challenges such as hardware stability, qubit coherence, and error correction continue to hinder the scalability of quantum computers~\cite{gill2024quantum, sood2024archives}.

Beyond its theoretical significance, QC is already demonstrating its potential impact across various industries. QAI is emerging as a transformative paradigm integrating quantum computing with machine learning to enhance model training, accelerate optimisation tasks, and improve data analysis capabilities~\cite{eswaran2024role, singh2024enhancing}. QAI applications extend across fields such as drug discovery, where quantum algorithms can expedite molecular simulations for pharmaceutical development~\cite{rawat2022quantum}, and materials science, where quantum-enhanced models can predict the properties of novel materials with unprecedented accuracy. Additionally, financial modelling and climate simulation are expected to benefit from quantum-enhanced data processing, leading to more accurate risk assessments and environmental predictions~\cite{krenn2023artificial}. The QC and AI synergy marks a significant shift in computational capabilities, reshaping both the technological landscape and its broader societal implications.

As research and development in QC progress, it is crucial to ensure that quantum algorithms—particularly those underpinning QAI—are developed with robustness, transparency, and ethical considerations in mind. Responsible QSE will play a pivotal role in guiding the ethical deployment of QAI, mitigating risks related to fairness, accountability, and accessibility as these technologies continue to evolve.

\section{Quantum Software Engineering and Ethical Dilemma} \label{QSE}

Quantum software is essential for realising the full potential of QC, including QAI. However, the fundamental principles of quantum mechanics, such as superposition and entanglement, introduce novel challenges for QSE~\cite{CACMQSE}, an emerging discipline focused on engineering dependable quantum software. Despite its significance, QSE is evolving without explicit consideration of its broader implications. Drawing from historical precedents in classical software engineering, we argue that QSE must be designed responsibly from the outset rather than retrofitted with ethical safeguards after its widespread adoption.

Quantum software operates fundamentally differently from classical software due to the unique characteristics of QC, including its inherently probabilistic nature, superposition, and entanglement~\cite{CACMQSE}. These features raise critical ethical and technical concerns, particularly in areas such as algorithmic governance, transparency, cryptography, and representational justice~\cite{QCProliferation}. 

First, the exponential computational power of QC enables the solution of complex problems previously deemed intractable on classical computers. While this unlocks new possibilities, it also carries significant geopolitical, privacy, and social implications~\cite{QCProliferation}. For instance, the ability of quantum algorithms to break classical encryption schemes threatens global cybersecurity, posing risks to financial institutions, medical records, and national security infrastructures.

Second, quantum software relies on computations performed in superposition, offering substantial computational advantages. However, when a quantum state is measured, it collapses into a definite classical state, making quantum computations inherently opaque. This black-box nature raises transparency and explainability concerns, particularly when quantum systems are deployed in high-stakes domains. A relevant example is quantum-enhanced fraud detection in banking, where quantum algorithms may flag suspicious transactions based on patterns that even experts cannot fully interpret, leading to unexplained account freezes or wrongful denials of financial services.

Third, the probabilistic nature of QC can introduce unpredictability in quantum software behaviour, an aspect absent in classical computing. This unpredictability complicates outcome verification and accountability, making it challenging to determine responsibility for decisions or errors arising from quantum systems~\cite{Possati2023}. Consider a quantum-powered predictive policing system that uses quantum-enhanced machine learning models to assess crime probabilities in urban areas. If such a system recommends increased surveillance in a particular neighbourhood due to quantum-driven probability estimates, yet those estimates cannot be independently verified, it may lead to unfair law enforcement practices and potential biases, reinforcing systemic discrimination.

Given these challenges, integrating responsible QSE principles into the foundational development of quantum software is imperative. Ethical considerations must be embedded in quantum software design to ensure accountability, fairness, and security, preventing the emergence of unintended risks as QAI and quantum computing technologies continue to evolve.

\textit{Responsible QSE} principles and practices must be developed in parallel with technical advancements rather than addressing ethical, security, and reliability concerns retrospectively. The history of classical AI has demonstrated that a lack of proactive governance can result in unintended algorithmic biases, security vulnerabilities, and ethical dilemmas~\cite{hevia2024quantum}. Lessons from AI governance emphasise the necessity of embedding \textbf{transparency, fairness, and accountability} into QSE from the outset~\cite{bano2024diversity}. This paper advocates for an interdisciplinary approach to QSE, integrating insights from \textit{quantum physics, software engineering, and AI ethics} to establish robust development frameworks and methodologies.

To advance responsible QSE, several key aspects require further investigation by the software engineering and QC communities:

\begin{itemize}
    \item \textbf{Quantum Software Verification and Debugging:} Unlike classical programs, quantum software does not produce deterministic outputs, making verification inherently challenging. \textit{Formal methods, quantum error correction (QEC), and hybrid verification techniques} are critical to ensuring software reliability~\cite{zhao2020quantum, ali2023quantum}. These approaches must integrate ethical considerations as first-class entities to mitigate risks associated with unpredictable quantum behaviour.  

    \item \textbf{Automated Quantum Software Engineering (AQSE):} As quantum hardware scales, manual software development becomes impractical. AQSE frameworks aim to automate quantum program synthesis, optimise quantum-classical interactions, and enhance usability through low-code and no-code platforms~\cite{sarkar2024automated}. However, ensuring that these frameworks are designed with \textit{ethics, transparency, explainability, and accountability} in mind is crucial to preventing the propagation of opaque or biased quantum applications.

    \item \textbf{Hybrid Classical-Quantum Development:} Most current quantum applications rely on hybrid classical-quantum systems, where classical pre- and post-processing is required for quantum computations. Quantum optimisation algorithms such as Quantum Approximate Optimisation Algorithms (QAOAs) exemplify this hybrid nature~\cite{QAOANewSurvey}. Developing \textit{efficient frameworks} to optimise \textit{data transfer, execution, and interoperability} remains a key research challenge~\cite{CACMQSE, ali2022software}. This raises critical ethical questions: Should classical ethical standards apply to quantum components, or do the unique properties of quantum computation necessitate a novel ethical framework?

    \item \textbf{Security and Quantum-Resilient Cryptography:} While QC threatens classical encryption systems, QSE must also address \textit{security vulnerabilities within quantum applications themselves}. The development of \textit{secure quantum software practices} and \textit{post-quantum cryptographic solutions} is imperative to safeguard sensitive data and prevent exploitation~\cite{QuantumSafe}.

    \item \textbf{Responsible and Ethical AI in Quantum Software:} As AI becomes increasingly integrated with QC, ethical safeguards are essential to prevent bias, discrimination, and opaque decision-making. AI models trained on quantum-enhanced datasets must uphold \textit{fairness and transparency}, ensuring that quantum-enhanced AI benefits a diverse range of users equitably. Addressing \textit{bias detection, fairness-aware algorithms, and equitable access to quantum resources} in QSE will be fundamental to maintaining ethical integrity in QAI applications~\cite{sarkar2024automated, Possati2023, perrier2021ethical}.

    \item \textbf{Algorithmic Fairness and Accountability:} Biases embedded in quantum datasets or models can lead to skewed outcomes, just as they have in classical AI. Ensuring \textit{bias detection and fairness-aware quantum algorithms} is crucial to prevent quantum systems from exacerbating existing societal inequalities~\cite{bano2024diversity, hevia2024quantum}. Future research must focus on developing techniques to audit quantum decision-making processes and establish frameworks for accountable quantum software engineering.
\end{itemize}

As QSE progresses, the need for global collaboration, ethical regulatory frameworks, and open-source quantum software initiatives will become increasingly urgent. This paper highlights perspectives on responsible QSE, aiming to raise awareness, encourage community discussions, and propose a roadmap for ethically aligned quantum software development.

\section{Ethics Related to the Convergence of Quantum Computing and AI} \label{sec:ethics}

As QC advances, its interplay with AI—both in terms of classical AI applied to QC, including QSE and QAI, must be examined through the lens of ethics, responsible governance, and inclusion~\cite{hoffmann2024business, umbrello2024ethics}. A key question is whether the ethical challenges posed by QC, QSE, and QAI are distinct enough to require entirely new frameworks, or whether existing AI ethics principles can be adapted to address quantum-specific risks. While classical AI ethics has made progress in promoting \textit{fairness, accountability, and transparency}, QAI introduces additional complexities that extend beyond these established concerns. The convergence of QC and AI demands an ethical paradigm considering the unique risks of quantum-enhanced computation, particularly its implications for decision-making, security, and global equity.

\subsection{Ethical Challenges} \label{subsec:ethicalchallenges}

The QC and AI convergence presents ethical dilemmas that extend beyond those of traditional computing technologies. QAI, in particular, raises concerns about algorithmic fairness and transparency, as its probabilistic nature could introduce biases that are more complex and harder to detect than those in classical AI~\cite{serebrenik2023diversity, rodriguez2021perceived}. If left unregulated, these biases could exacerbate social inequalities in applications such as hiring algorithms, predictive policing, and credit scoring~\cite{bano2024vision, shams2023ai}. To address these risks, ethical AI in the quantum era must incorporate bias detection mechanisms specifically designed for quantum systems, ensuring that quantum-powered decisions remain interpretable and justifiable.

\subsection{Responsible and Ethical QSE and AI} \label{subsec:ResponsibleQSEandAI}

Responsible AI development within QSE must prioritise ethical considerations such as fairness, transparency, and accountability. QAI models built using quantum-enhanced datasets must be designed to prevent the reinforcement of discrimination and bias, particularly in high-stakes applications like healthcare, law enforcement, and finance~\cite{bano2025does, sarkar2023automated}. Given the inherent complexity of quantum computations, ensuring that QAI remains explainable and auditable is crucial, as quantum logic is significantly less interpretable than classical approaches~\cite{bano2024vision, shams2023ai}.

Diversity and inclusion (D\&I) also play a vital role in responsible QSE and QAI development. The ongoing lack of diversity in classical AI and software engineering has already hindered innovation and contributed to biased outcomes for underrepresented communities~\cite{shams2023ai, albusays2021diversity}. In QSE and QAI, D\&I must extend beyond demographic representation to include diverse perspectives in algorithmic design, data processing, and governance structures~\cite{Zowghi2023DiversityAI}. A more inclusive quantum software development process will lead to better-designed quantum algorithms that address the needs of a broader user base while mitigating systemic biases~\cite{IntegratingEDI}.

Accessibility remains a major challenge in QSE and QAI, with concerns over monopolistic control of quantum resources. Without proactive regulation, access to quantum infrastructure could remain concentrated within a few well-funded nations and corporations, exacerbating global technological divides~\cite{seskir2023democratization}. To prevent digital colonisation and ensure that quantum advancements serve humanity as a whole, ethical frameworks for QSE must prioritise open access and equitable distribution of quantum resources~\cite{meyer2024disparities}.

\subsection{Developing a Unified Ethical Framework for QSE and QAI} \label{subsec:unifiedfarmework}

Existing AI ethical frameworks must evolve to address the complexities introduced by QC. Current guidelines primarily focus on algorithmic bias and transparency while assuming the security of data encryption methods~\cite{umbrello2024ethics}. However, quantum advancements could render many of these principles inadequate, particularly in areas such as data security, consent, and algorithmic accountability. Ethical governance frameworks must be adapted to account for the probabilistic nature of quantum computations and the unprecedented scale at which QAI processes information.

A crucial step in this evolution is encouraging international collaboration on ethical QAI standards. Ethical governance structures for QC should not be dictated by a single nation or corporation but should be developed collectively, incorporating diverse perspectives from academia, industry, and regulatory bodies~\cite{RQCPrinciple}. Similar to nuclear non-proliferation agreements, a global commitment to responsible quantum development could help mitigate the risks of unchecked technological dominance.

Without coordinated oversight, the rapid advancement of QAI may outpace regulatory efforts, leading to unintended consequences. Governments, researchers, and private sector stakeholders must work together to establish policies that uphold transparency, accountability, and fairness in QAI applications. By embedding ethical considerations into QSE from the outset, we can guide the responsible development of quantum-powered AI while preventing its monopolisation and potential misuse.

\subsection{The Known-Unknowns of the Convergence of QC and AI}

QC and AI are advancing rapidly, yet their long-term impact remains uncertain \cite{de2022own}. While QAI promises breakthroughs in optimisation, security, and decision-making, it also introduces profound ethical and governance dilemmas. Who will control these technologies? How will they shape economies, societies, and digital sovereignty? These \textit{known-unknowns} demand urgent discussion, as their answers will define the future of responsible QC, including QSE and AI. Below, we outline key questions that require exploration and resolution:

\begin{itemize}
    \item \textbf{Governance and Power Structures:} QC has the potential to concentrate power among the first nations and corporations to develop it~\cite{johnson2019governance, csenkey2023post, purohit2024building, gasser2024call}. Will early adopters dictate quantum encryption standards, AI governance, and global digital infrastructure? If access to QC remains restricted, could this lead to \textit{quantum colonialism}, where quantum-capable entities dominate technologically dependent regions? The opacity of quantum algorithms further complicates governance—if QAI decisions in healthcare, finance, or criminal justice cannot be fully explained or audited, should they be trusted? If these systems surpass human oversight, who bears responsibility for their failures?

    \item \textbf{Equity and Bias in Quantum AI:} Despite its transformative potential, QC remains inaccessible to many due to high costs and the need for specialised expertise \cite{AIQCConvergence, chauhan2025applications}. If this exclusivity persists, will quantum breakthroughs reinforce existing disparities, benefiting only a select few? Classical AI has already exhibited biases in hiring, policing, and financial lending—how can we prevent QAI from amplifying these biases exponentially? As quantum algorithms process increasingly complex datasets, traditional fairness metrics may be insufficient. How do we ensure that QAI serves all, rather than deepening systemic discrimination?

    \item \textbf{Security Risks and Ethical Trade-offs:} The ability of quantum computers to break classical encryption threatens global cybersecurity~\cite{singh2024enhancing, paul2025integration, rawat2022quantum}. Should governments and industries transition to quantum-safe encryption now, despite the logistical and financial challenges? Beyond cryptography, QC could enable mass surveillance by decrypting private communications in real-time. Would such capabilities be justifiable in the name of national security, or would they mark an irreversible erosion of civil liberties? With quantum technologies holding both constructive and destructive potential, how do we balance innovation with ethical responsibility?

    \item \textbf{The Uncharted Future of Quantum AI:} As QAI advances, it may outpace regulatory capabilities, reshaping governance and decision-making~\cite{de2022own, RQCPrinciple, taylor2020quantum}. Will human-driven governance remain relevant if quantum algorithms can model economic, political, and societal shifts with near-perfect accuracy? The rise of quantum-generative AI could create hyper-realistic deepfakes and misinformation. How will we maintain trust in digital content when fabricated realities become indistinguishable from truth? Who should regulate these systems, and can ethical oversight keep pace with quantum acceleration?

\end{itemize}

\textbf{A Defining Moment:} The future of QC convergence, including QSE and AI, is being shaped now~\cite{pooranam2023quantum, AIQCConvergence, weigang2022new}. Unlike previous technological revolutions, where ethics were considered only after harm was done, quantum technologies demand proactive governance. Policymakers, researchers, and industry leaders must confront these known-unknowns today, ensuring that transparency, fairness, and accountability are embedded in QAI before it is too late. Will QC be an instrument of empowerment or exclusion? The trajectory of quantum technology—and its role in shaping our world—depends on the decisions we make now.

\section{Concluding Remarks}

Quantum computing is the precipice of redefining technological power, economic structures, and global governance. Whether this transformation serves as a force for equitable progress or exacerbates existing societal divides depends on our choices today. The rapid evolution of quantum technologies presents both an unprecedented opportunity and an urgent ethical challenge. If left unregulated, QC risks entrenching disparities, concentrating power among a privileged few, and reinforcing systemic biases in AI and decision-making systems. However, if approached with foresight, inclusion, and ethical responsibility, QC has the potential to enhance innovation that benefits humanity as a whole. 

Technology's history is littered with missed opportunities to embed fairness and inclusivity from the outset. The evolution of classical AI and software engineering has demonstrated the dangers of overlooking ethical considerations until biases become ingrained in systems that shape our daily lives. Quantum software engineering must learn from these failures. The ethical trajectory of QC should not be a retrospective fix but an intentional, proactive effort to integrate fairness, transparency, and accountability into quantum software development from its inception. This requires a collective commitment from researchers, policymakers, and industry leaders to ensure that QC does not simply reflect the inequalities of the past but becomes a tool for addressing them. 

To achieve this vision, concrete action is required. The research community must prioritise frameworks that embed ethical considerations into quantum algorithm design, security models, and AI governance. Policymakers must anticipate the risks of monopolisation and implement regulations that promote equitable access to quantum resources. Industry leaders must commit to developing quantum applications that prioritise societal benefit over short-term competitive advantage. A global research agenda that integrates responsible AI principles, bias mitigation, and diversity into QSE is crucial to shaping the future of QC in a way that empowers rather than excludes. 

The future of quantum computing is not preordained—it is a construct of today's ethical and technical decisions. As we stand on the threshold of a quantum revolution, we must ask ourselves: will this technology amplify existing power hierarchies, or will it be harnessed to create a more just, inclusive, and transparent digital world? The window for shaping ethical quantum software engineering is now. It is not merely an academic challenge, but a moral imperative—one that will define the legacy of quantum computing for generations to come.

\bibliographystyle{unsrt}  
\bibliography{references}  

\begin{thebibliography}{10}

\bibitem{preskill2018quantum}
John Preskill.
\newblock Quantum computing in the nisq era and beyond.
\newblock {\em Quantum}, 2:79, 2018.

\bibitem{schuld2019quantum}
Maria Schuld and Nathan Killoran.
\newblock Quantum machine learning in feature hilbert spaces.
\newblock {\em Physical review letters}, 122(4):040504, 2019.

\bibitem{wang2023recent}
Pei-Hua Wang, Jen-Hao Chen, Yu-Yuan Yang, Chien Lee, and Yufeng~Jane Tseng.
\newblock Recent advances in quantum computing for drug discovery and development.
\newblock {\em IEEE Nanotechnology Magazine}, 17(2):26--30, 2023.

\bibitem{herman2023quantum}
Dylan Herman, Cody Googin, Xiaoyuan Liu, Yue Sun, Alexey Galda, Ilya Safro, Marco Pistoia, and Yuri Alexeev.
\newblock Quantum computing for finance.
\newblock {\em Nature Reviews Physics}, 5(8):450--465, 2023.

\bibitem{ho2024quantum}
Kin Tung~Michael Ho, Kuan-Cheng Chen, Lily Lee, Felix Burt, Shang Yu, and Po-Heng Lee.
\newblock Quantum computing for climate resilience and sustainability challenges.
\newblock In {\em 2024 IEEE International Conference on Quantum Computing and Engineering (QCE)}, volume~2, pages 262--267. IEEE, 2024.

\bibitem{ali2022software}
Shaukat Ali, Tao Yue, and Rui Abreu.
\newblock When software engineering meets quantum computing.
\newblock {\em Communications of the ACM}, 65(4):84--88, 2022.

\bibitem{de2024quantum}
Manuel De~Stefano, Fabiano Pecorelli, Dario Di~Nucci, Fabio Palomba, and Andrea De~Lucia.
\newblock The quantum frontier of software engineering: A systematic mapping study.
\newblock {\em Information and Software Technology}, page 107525, 2024.

\bibitem{sarkar2024automated}
Aritra Sarkar.
\newblock Automated quantum software engineering.
\newblock {\em Automated Software Engineering}, 31(1):1--17, 2024.

\bibitem{hoffmann2024business}
Christian~Hugo Hoffmann and Frederik~F. Flother.
\newblock Why business adoption of quantum and ai technology must be ethical.
\newblock {\em Research Directions: Quantum Technologies}, 2:e4, 2024.

\bibitem{shams2023ai}
Rifat~Ara Shams, Didar Zowghi, and Muneera Bano.
\newblock {AI} and the quest for diversity and inclusion: A systematic literature review.
\newblock {\em AI and Ethics}, pages 1--28, 2023.

\bibitem{bano2024diversity}
Muneera Bano, Didar Zowghi, Fernando Mourao, Sarah Kaur, and Tao Zhang.
\newblock Diversity and inclusion in ai for recruitment: Lessons from industry workshop.
\newblock {\em arXiv preprint arXiv:2411.06066}, 2024.

\bibitem{bano2025does}
Muneera Bano, Hashini Gunatilake, and Rashina Hoda.
\newblock What does a software engineer look like? exploring societal stereotypes in llms.
\newblock {\em arXiv preprint arXiv:2501.03569}, 2025.

\bibitem{AIQCConvergence}
G~Viggiano and D~Brin.
\newblock {\em Convergence: Artificial Intelligence and Quantum Computing: Social, Economic, and Policy Impacts}.
\newblock Wiley, 2023.

\bibitem{seskir2023democratization}
Zeki~C Seskir, Steven Umbrello, Christopher Coenen, and Pieter~E Vermaas.
\newblock Democratization of quantum technologies.
\newblock {\em Quantum Science and Technology}, 8(2):024005, 2023.

\bibitem{ten2023reading}
Carolyn Ten~Holter, Philip Inglesant, and Marina Jirotka.
\newblock Reading the road: challenges and opportunities on the path to responsible innovation in quantum computing.
\newblock {\em Technology Analysis \& Strategic Management}, 35(7):844--856, 2023.

\bibitem{bcg2022quantumrace}
{Boston Consulting Group}.
\newblock The u.s., china, and europe are ramping up a quantum computing arms race.
\newblock {\em Boston Consulting Group}, September 2022.
\newblock Accessed: 2024-10-14.

\bibitem{meyer2023question}
Josephine~C Meyer, Gina Passante, and Bethany~R Wilcox.
\newblock The question of equity: Who has access to us quantum information education programs?
\newblock {\em arXiv preprint arXiv:2309.08629}, 2023.

\bibitem{meyer2024disparities}
Josephine~C Meyer, Gina Passante, and Bethany Wilcox.
\newblock Disparities in access to us quantum information education.
\newblock {\em Physical Review Physics Education Research}, 20(1):010131, 2024.

\bibitem{shor2002introduction}
Peter~W Shor.
\newblock Introduction to quantum algorithms.
\newblock In {\em Proceedings of Symposia in Applied Mathematics}, volume~58, pages 143--160, 2002.

\bibitem{rungta2009quadratic}
Pranaw Rungta.
\newblock The quadratic speedup in grover's search algorithm from the entanglement perspective.
\newblock {\em Physics Letters A}, 373(31):2652--2659, 2009.

\bibitem{gill2024quantum}
Sukhpal~Singh Gill, Oktay Cetinkaya, Stefano Marrone, Elias~F Combarro, Daniel Claudino, David Haunschild, Leon Schlote, Huaming Wu, Carlo Ottaviani, Xiaoyuan Liu, et~al.
\newblock Quantum computing: Vision and challenges.
\newblock {\em arXiv preprint arXiv:2403.02240}, 2024.

\bibitem{sood2024archives}
Vaishali Sood and Rishi~Pal Chauhan.
\newblock Archives of quantum computing: research progress and challenges.
\newblock {\em Archives of Computational Methods in Engineering}, 31(1):73--91, 2024.

\bibitem{eswaran2024role}
Ushaa Eswaran, Alex Khang, and Vishal Eswaran.
\newblock Role of quantum computing in the era of artificial intelligence (ai).
\newblock In {\em Applications and Principles of Quantum Computing}, pages 46--68. IGI Global, 2024.

\bibitem{singh2024enhancing}
Shoumya Singh and Deepak Kumar.
\newblock Enhancing cyber security using quantum computing and artificial intelligence: A review.
\newblock {\em algorithms}, 4(3), 2024.

\bibitem{rawat2022quantum}
Bhupesh Rawat, Nidhi Mehra, Ankur~Singh Bist, Muhamad Yusup, and Yulia Putri~Ayu Sanjaya.
\newblock Quantum computing and ai: Impacts \& possibilities.
\newblock {\em ADI Journal on Recent Innovation}, 3(2):202--207, 2022.

\bibitem{krenn2023artificial}
Mario Krenn, Jonas Landgraf, Thomas Foesel, and Florian Marquardt.
\newblock Artificial intelligence and machine learning for quantum technologies.
\newblock {\em Physical Review A}, 107(1):010101, 2023.

\bibitem{CACMQSE}
Shaukat Ali, Tao Yue, and Rui Abreu.
\newblock When software engineering meets quantum computing.
\newblock {\em Commun. ACM}, 65(4):84–88, mar 2022.

\bibitem{QCProliferation}
Dominic Rosch-Grace and Jeremy Straub.
\newblock Analysis of the likelihood of quantum computing proliferation.
\newblock {\em Technology in Society}, 68:101880, 2022.

\bibitem{Possati2023}
Luca~M. Possati.
\newblock Ethics of quantum computing: an outline.
\newblock {\em Philosophy {\&} Technology}, 36(3):48, Jul 2023.

\bibitem{hevia2024quantum}
Jose~Luis Hevia, Guido Peterssen, and Mario Poattini.
\newblock Quantum software development risks.
\newblock {\em Quantum Information and Computation}, 24(5\&6):0455--0467, 2024.

\bibitem{zhao2020quantum}
Jianjun Zhao.
\newblock Quantum software engineering: Landscapes and horizons.
\newblock {\em CoRR}, abs/2007.07047, 2020.

\bibitem{ali2023quantum}
Shaukat Ali and Tao Yue.
\newblock Quantum software testing: A brief introduction.
\newblock In {\em 2023 IEEE/ACM 45th International Conference on Software Engineering: Companion Proceedings (ICSE-Companion)}, pages 332--333. IEEE, 2023.

\bibitem{QAOANewSurvey}
Kostas Blekos, Dean Brand, Andrea Ceschini, Chiao-Hui Chou, Rui-Hao Li, Komal Pandya, and Alessandro Summer.
\newblock A review on quantum approximate optimization algorithm and its variants.
\newblock {\em Physics Reports}, 1068:1--66, 2024.

\bibitem{QuantumSafe}
Lei Zhang, Andriy Miranskyy, Walid Rjaibi, Greg Stager, Michael Gray, and John Peck.
\newblock Making existing software quantum safe: A case study on ibm db2.
\newblock {\em Information and Software Technology}, 161:107249, 2023.

\bibitem{perrier2021ethical}
Elija Perrier.
\newblock Ethical quantum computing: A roadmap.
\newblock {\em arXiv preprint arXiv:2102.00759}, 2021.

\bibitem{umbrello2024ethics}
Steven Umbrello.
\newblock Ethics of quantum technologies: A scoping review.
\newblock {\em International Journal of Applied Philosophy}, 2024.

\bibitem{serebrenik2023diversity}
Alexander Serebrenik.
\newblock Diversity and inclusion in software engineering.
\newblock {\em Software Engineering 2023 Fachtagung des GI-Fachbereichs Softwaretechnik}, page~21.

\bibitem{rodriguez2021perceived}
Gema Rodr{\'\i}guez-P{\'e}rez, Reza Nadri, and Meiyappan Nagappan.
\newblock Perceived diversity in software engineering: a systematic literature review.
\newblock {\em Empirical Software Engineering}, 26:1--38, 2021.

\bibitem{bano2024vision}
Muneera Bano, Didar Zowghi, and Vincenzo Gervasi.
\newblock A vision for operationalising diversity and inclusion in {AI}.
\newblock In {\em Proceedings of the 2nd International Workshop on Responsible AI Engineering}, pages 36--45, 2024.

\bibitem{sarkar2023automated}
Aritra Sarkar.
\newblock Automated quantum software engineering.
\newblock {\em Automated Software Engineering}, 31(1):36, Apr 2024.

\bibitem{albusays2021diversity}
Khaled Albusays, Pernille Bjorn, Laura Dabbish, Denae Ford, Emerson Murphy-Hill, Alexander Serebrenik, and Margaret-Anne Storey.
\newblock The diversity crisis in software development.
\newblock {\em IEEE Software}, 38(2):19--25, 2021.

\bibitem{Zowghi2023DiversityAI}
Didar Zowghi and Francesca da~Rimini.
\newblock Diversity and inclusion in artificial intelligence.
\newblock {\em ArXiv}, abs/2305.12728, 2023.

\bibitem{IntegratingEDI}
Milka Nyariro, Elham Emami, and Samira Abbasgholizadeh~Rahimi.
\newblock Integrating equity, diversity, and inclusion throughout the lifecycle of artificial intelligence in health.
\newblock In {\em 13th Augmented Human International Conference}, AH2022, New York, NY, USA, 2022. Association for Computing Machinery.

\bibitem{RQCPrinciple}
Mauritz Kop, Mateo Aboy, Eline Jong, Urs Gasser, Timo Minssen, I.~Cohen, Mark Brongersma, Teresa Quintel, Luciano Floridi, and Ray Laflamme.
\newblock 10 principles for responsible quantum innovation.
\newblock {\em SSRN Electronic Journal}, 01 2023.

\bibitem{de2022own}
Eline de~Jong.
\newblock Own the unknown: an anticipatory approach to prepare society for the quantum age.
\newblock {\em Digital Society}, 1(2):15, 2022.

\bibitem{johnson2019governance}
Walter~G Johnson.
\newblock Governance tools for the second quantum revolution.
\newblock {\em Jurimetrics}, 59(4):487--522, 2019.

\bibitem{csenkey2023post}
Kristen Csenkey and Nina Bindel.
\newblock Post-quantum cryptographic assemblages and the governance of the quantum threat.
\newblock {\em Journal of Cybersecurity}, 9(1):tyad001, 2023.

\bibitem{purohit2024building}
Abhishek Purohit, Maninder Kaur, Zeki~Can Seskir, Matthew~T Posner, and Araceli Venegas-Gomez.
\newblock Building a quantum-ready ecosystem.
\newblock {\em IET Quantum Communication}, 5(1):1--18, 2024.

\bibitem{gasser2024call}
Urs Gasser, Eline De~Jong, and Mauritz Kop.
\newblock A call for responsible quantum technology.
\newblock {\em Nature Physics}, 20(4):525--527, 2024.

\bibitem{chauhan2025applications}
Dipti Chauhan, Pragya Ranka, Pritika Bahad, and Rupali Pathak.
\newblock Applications of quantum artificial intelligence: A systematic review.
\newblock {\em Integration of AI, Quantum Computing, and Semiconductor Technology}, pages 159--182, 2025.

\bibitem{paul2025integration}
Shyamalendu Paul, Nobhonil~Roy Choudhury, Bipradash Pandit, and Avrodeep Dawn.
\newblock Integration of ai and quantum computing in cybersecurity: A comprehensive review.
\newblock {\em Integration of AI, Quantum Computing, and Semiconductor Technology}, pages 287--308, 2025.

\bibitem{taylor2020quantum}
Richard~D Taylor.
\newblock Quantum artificial intelligence: a “precautionary” us approach?
\newblock {\em Telecommunications Policy}, 44(6):101909, 2020.

\bibitem{pooranam2023quantum}
N~Pooranam, D~Surendran, N~Karthikeyan, and G~Ignisha Rajathi.
\newblock Quantum computing: future of artificial intelligence and its applications.
\newblock {\em Quantum Computing and Artificial Intelligence: Training Machine and Deep Learning Algorithms on Quantum Computers}, 163, 2023.

\bibitem{weigang2022new}
Li~Weigang, Liriam~Michi Enamoto, Denise~Leyi Li, and Geraldo~Pereira Rocha~Filho.
\newblock New directions for artificial intelligence: human, machine, biological, and quantum intelligence.
\newblock {\em Frontiers of Information Technology \& Electronic Engineering}, 23(6):984--990, 2022.

\end{thebibliography}






\end{document}